# Observation of robust corner states in photonic crystals without global symmetries

Zitao Ji[1,2,#], Jianfeng Chen[3,#,*], Yidong Zheng[1], Zhaogang Dong[2,*], and Zhi-Yuan Li[1,*]

[1] School of Physics and Optoelectronics, South China University of Technology, Guangzhou 510640, China

[2] Science, Mathematics, and Technology (SMT), Singapore University of Technology and Design, 8 Somapah Road, 487372, Singapore

[3] State Key Laboratory for Mesoscopic Physics and Frontiers Science Center for Nano-optoelectronics, School of Physics, Peking University, Beijing 100871, China

[*] Correspondence: phzyli@scut.edu.cn; cjf@pku.edu.cn; zhaogang_dong@sutd.edu.sg

[#] These authors contributed equally to this work.

**Abstract:** Robust corner states have attracted considerable attention in recent years, particularly in photonic systems. They are typically realized in second-order topological insulators, including shrunken-expanded lattices and breathing Kagome lattices, where global symmetries are essential in their formation. This reliance on symmetry, however, can make these states highly sensitive to structural disorder. Here, we propose and experimentally demonstrate a class of robust corner states in photonic crystals that require neither global symmetry nor nontrivial topology. By locally tailoring the corner geometry of an otherwise topologically trivial structure, we create strongly localized modes that remain stable under both position and radius disorder. These states are not accidental; rather they originate from a pair of Dirac points in the bulk band structure. Geometric tuning shifts the flatband edge states connecting the Dirac points into an otherwise trivial bandgap, thereby inducing strongly localized corner states. Our findings uncover a previously overlooked class of symmetry-independent, disorder-resilient corner states, extend robust localization beyond conventional topological protection, and provide a route toward light localization in all-dielectric photonic systems.

## Introduction

Robust corner states have emerged as a hallmark of second-order topological insulators, which generalize the conventional bulk-boundary correspondence by supporting boundary modes of codimension two[1–10]. In two-dimensional second-order topological insulators, these higher-codimension modes appear as zero-dimensional corner states[11–16]. Originally conceived for electronic systems, such second-order topological phases have since been extended to a broad range of other physical platforms, including photonic systems[3,17–29], acoustic systems[30–32], thermal metamaterials[33–35], electrical circuits[36–38], and ultracold atoms[39]. In photonic systems, two widely studied design paradigms have been employed to realize robust corner states. The first employs shrunken-expanded lattices, in which lattice deformations drive band inversions and alter quantized bulk multipole moments or edge polarizations, as characterized by Zak phases or related crystalline topological indices[3,22–24,29]. The second employs breathing Kagome lattices, in which corner states are associated with an obstructed atomic limit characterized by a mismatch between the Wannier centers and the physical lattice sites[25–28]. These zero-dimensional corner-localized modes offer promising opportunities for applications including topological lasers[40–45] and rainbow trapping[46,47], highlighting their versatility and potential for integrated photonics.

The conventional understanding is that robust corner states in higher-order topological phases generally rely on global crystalline symmetries, which quantize the relevant bulk or edge topological indices and constrain the corner modes. This reliance can make such states particularly sensitive to structural disorder that breaks the protecting symmetries. Rather than being immediately destroyed by any symmetry-breaking perturbation, their eigenfrequencies may shift, and their localization may eventually be lost through hybridization with edge or bulk continua. This vulnerability contrasts with that of first-order topological boundary states in systems such as Chern and $\mathbb{Z}_2$ insulators, which can survive substantial disorder provided that the bulk or mobility gap remains open and, in the $\mathbb{Z}_2$ case, time-reversal symmetry is preserved. This contrast raises a more fundamental question: must robust corner localization necessarily rely on global crystalline-symmetry protection? Recent theoretical work has shown that non-Hermiticity can provide an intrinsic mechanism for generating localized zero modes without global crystalline-symmetry protection[48]. In such systems, however, localization relies on explicitly non-Hermitian ingredients, such as asymmetric gain/loss or nonreciprocal coupling. Whether analogous symmetry-independent corner localization can be achieved in a fully Hermitian system, without relying on gain/loss, nonreciprocity, or other non-Hermitian effects, remains largely unexplored.

Here, we demonstrate a route to robust corner localization in Hermitian systems that is independent of global crystalline-symmetry protection. Our key finding is that the flatband edge states connecting the boundary projections of a pair of bulk Dirac points can be spectrally shifted through boundary engineering. Conventionally, corner states are obtained by breaking lattice symmetry or modulating intercell couplings, thereby lifting the Dirac-point degeneracy and opening a topological nontrivial bandgap (Figs. 1a-1b). Once the gap opens, the pre-existing flatband edge states persist within the newly opened gap, providing the in-gap boundary states from which corner states emerge. Opening a gap at the Dirac points, however, is not required. Instead, boundary engineering can shift these Dirac-connected flatband edge states into a separate, topologically trivial bandgap, thereby generating robust edge and corner states (Fig. 1c). Tight-binding calculations reveal that tuning an on-site mass term $M_s$ at the boundary of a Dirac semimetal shifts these flatband edge states from their original spectral region between the projected Dirac points, through the upper bulk band, and into the trivial bandgap (Figs. 1d-1e). Joining two such engineered boundaries to form a corner then gives rise to a corner-localized state (Figs. 1f-1g). The resulting corner states evolve from the Dirac-connected edge states rather than from isolated defect modes and require neither higher-order bulk topology nor global crystalline-symmetry protection[1,11,49–52]. A detailed comparison is provided in Table S1 of Supplementary Note 4. Notably, in electronic systems, these flatband edge states may lie far from the Fermi level, limiting their direct contribution to low-energy responses. This constraint does not apply to photonic crystals, which are not restricted by a Fermi level. Detail tight-binding calculations are presented in Supplementary Note 1.

We experimentally validate this mechanism in a two-dimensional photonic crystal by locally tailoring its boundary geometry, without globally deforming the lattice or invoking a higher-order bulk topological invariant. Near-field measurements directly reveal the corner-localized modes, which retain both their spectral isolation and spatial localization under positional and radius disorder. In contrast to conventional symmetry-dependent corner states protected by global lattice symmetries, the states reported here preserve their localization under local structural perturbations, whereas symmetry-protected corner states undergo rapid bandgap closure and lose corner localization under comparable disorder.

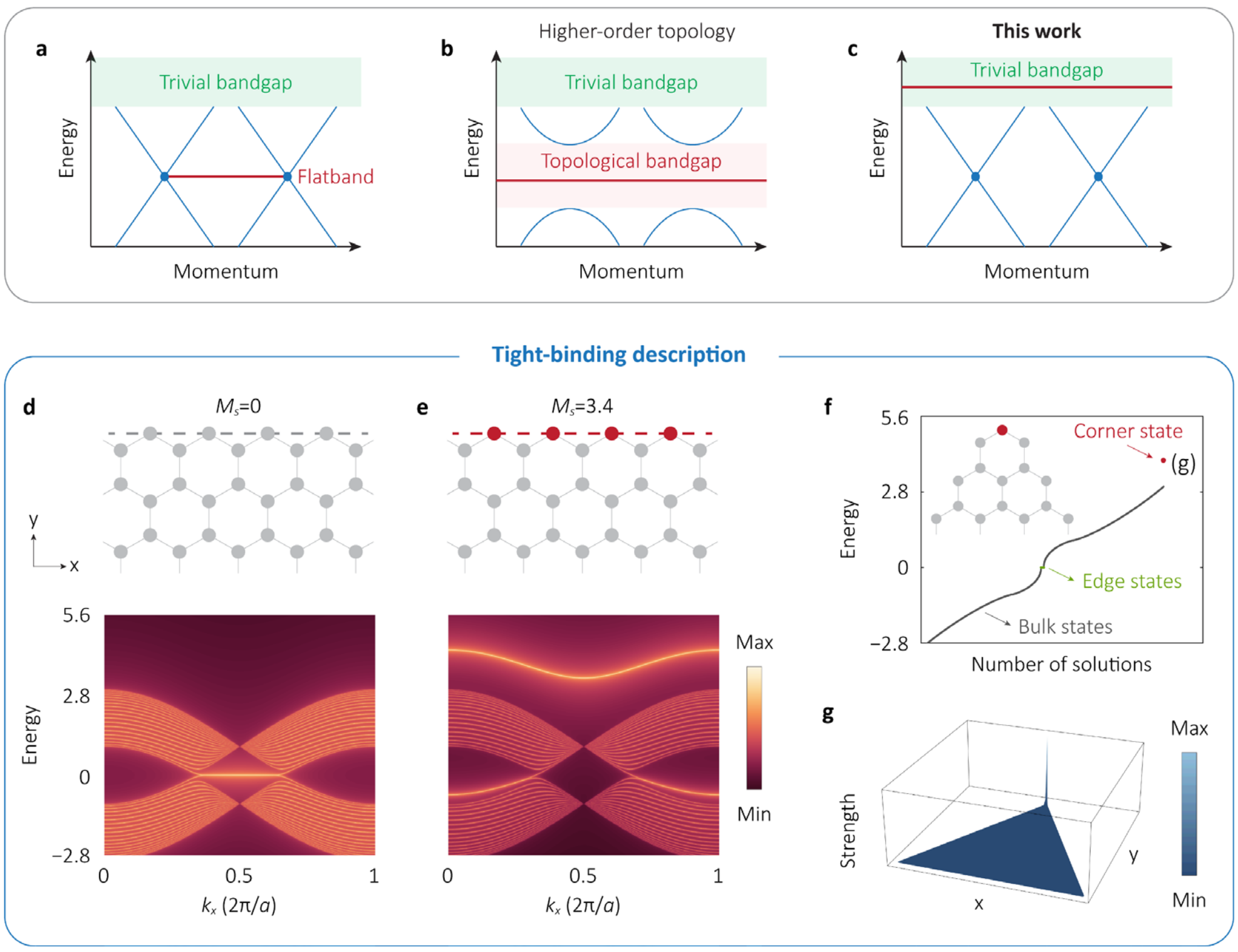

**Fig. 1. Robust corner states derived from Dirac-connected flatband edge states.** (a-c) Two routes to creating edge and corner states from the flatband edge states of a Dirac semimetal. (a) Projected band structure of a Dirac semimetal, showing flatband edge states connecting the boundary projections of a pair of bulk Dirac points. (b) Lifting the Dirac-point degeneracy opens a topologically nontrivial bandgap, within which the Dirac-connected edge states persist and give rise to topological boundary states. (c) Without lifting the Dirac-point degeneracy, relocating the flatband edge states into a separate, topologically trivial bandgap also produces spectrally isolated edge and corner states. (d-g) Tight-binding description of the proposed mechanism. (d) Lattice geometry and projected band structure of the Dirac semimetal; the color scale denotes the surface

density of states. (e) Boundary engineering shifts the Dirac-connected flatband edge states through the upper bulk band and into a trivial bandgap. (f) Finite rhombic lattice and its eigenvalue spectrum. (g) Spatial field distribution of the corner-state eigenmode.

## Results

**Photonic design.** To engineer flatband edge states within a topologically trivial bandgap, we start from a two-dimensional honeycomb lattice with a zigzag termination (Figs. 2a-2c). The structure comprises dielectric cylinders with lattice constant $a$, radius $r$=0.176$a$, and relative permittivity $\varepsilon$=11.5. The dielectric profile is translationally invariant along the $z$ direction, allowing the eigenmodes to be separated into transverse magnetic and transverse electric polarizations. In this work, we focus on transverse magnetic modes; transverse electric modes can be analyzed analogously. To visualize how the flatband edge state is displaced across the upper bulk bands and eventually relocated into a trivial bandgap, we calculate the surface density of states of semi-infinite photonic crystals with open boundaries (Figs. 2a-2c)[53]. When all cylinders have identical radii and the structure is unperturbed, the flatband edge state lies between two Dirac degeneracies at the K and K' points (Fig. 2a), in agreement with the tight-binding result (Fig. 1d). The slight bending of the flatband edge state originates from the intrinsic dispersion of the photonic crystal. In a Dirac semimetal system, a change in the boundary on-site mass can be readily implemented in photonic crystals by modifying the size of the boundary cylinders. Accordingly, we tailor only the outermost row of cylinders by reducing the boundary radius to $r_E$ = 0.8$r$, as indicated by the white dashed line in Fig. 2b. As a result, the flatband edge state shifts upward into the bulk continuum. Upon further decreasing the boundary radius to $r_E$ = 0.50$r$ (Fig. 2c), the flatband edge state traverses the upper bulk band and becomes fully isolated within the trivial bandgap. Additional surface density of states for different $r_E$ are shown in Supplementary Note 2.

**Programmable corner states.** The isolation of the flatband edge state within a trivial bandgap, together with its strong frequency tunability enabled by local boundary tailoring, provides a direct and flexible route to realizing highly programmable corner states. This mechanism allows deterministic control over both the number of corner modes and their spectral positions within the bandgap, without altering the global lattice geometry or bulk topology. We independently introduce three corner modes in a triangular photonic crystal by locally tuning the radius ratio $r_E/r$ of the cylinders located at the top, left, and right vertices (Fig. 2d). Specifically, the boundary radii of these three corner cylinders are set to $r_E$ = 0.5$r$, 0.4$r$, and 0.33$r$, respectively. Figure 2e presents the eigenfrequency spectrum of the designed structure, where bulk and corner-localized modes are clearly distinguished. Each corner mode emerges at a distinct frequency within the bandgap, with normalized eigenfrequencies of 0.413, 0.446, and 0.475 corresponding to the left, top, and right corners, respectively. It should be noted that edge states associated with the finite lattice are not involved in the reported corner localization, since only the corner units are modified, while the bulk and edges remain unchanged. Figures 2f-2h display the electric field intensity distributions of these corner-localized modes under frequency-resolved excitation, confirming their strong spatial confinement. The simulation configuration is illustrated in Supplementary Note 3. These results demonstrate broadband spectral tunability of corner states achieved solely through local geometry modification. Crucially, this approach enables the formation of corner states without invoking global symmetry breaking or topological bandgap engineering. Instead, the corner states originate from a flatband edge state inherited from paired Dirac points and are selectively positioned inside a trivial bandgap through local boundary control. We also reveal that compressing the perfect electric conductor boundaries can also realize corner states in the trivial bandgap (see Supplementary Note 4).

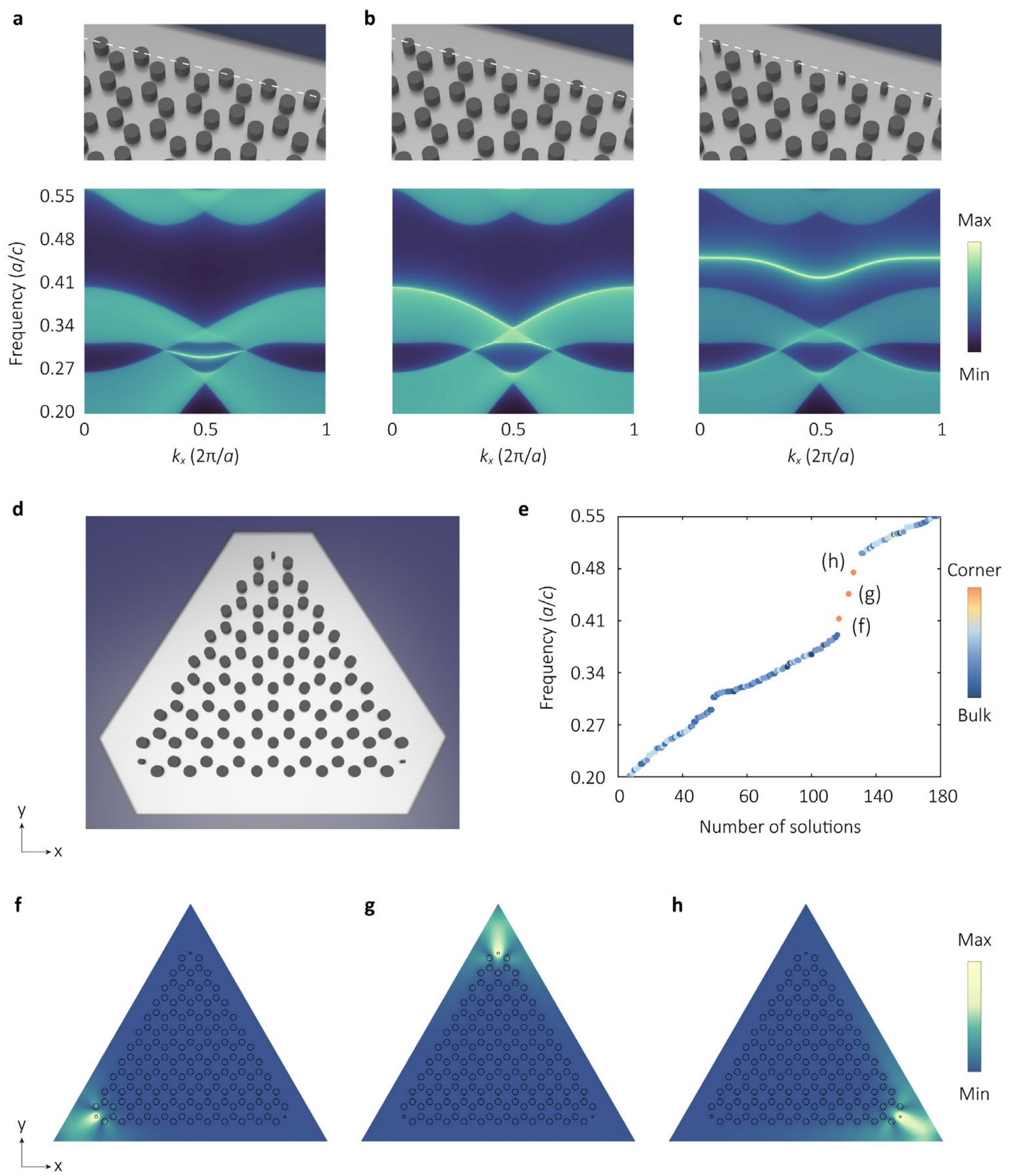

**Fig. 2. Programmable corner states in photonic crystals.** (a-c) Close-up views of the sample geometry and the corresponding projected bandstructures; the color scale denotes the surface density of states. The radius of the bulk is $r$ = 0.176$a$, while the radius of the boundary cylinders (white dashed lines) is $r_E = r$ in (a), $r_E = 0.8r$ in (b), and $r_E = 0.5r$ in (c), respectively. (d) Schematic of a triangular photonic crystal placed on an aluminum plate. The surrounding blue regions denote absorbing materials used to suppress unwanted reflections. (e) Eigenfrequency spectrum of the triangular photonic crystal, where bulk and corner states are distinguished. (f-h) Eigenmodal field distributions of the corner states, which appear at different frequencies due to the unequal cylinder radii at the three corners.

**Experimental identification.** We fabricate photonic crystal samples to experimentally observe the corner states and benchmark the results against numerical simulations. The sample is a triangular photonic crystal with lattice constant $a = 17$ mm, consisting of 11 units along each side. The structure is composed of nonmagnetized yttrium iron garnet (YIG) cylinders with relative dielectric permittivity $\varepsilon = 11.5$, relative permeability $\mu = 1$, and a radius $r = 3$ mm. To suppress radiation leakage into free space, the photonic crystal is sandwiched between two flat aluminum plates, which act as effective perfect electric conductor boundaries. The structure is excited by a broadband quasi-plane wave generated by a horn antenna, while the out-of-plane electric field $E_z$ is detected using a probe antenna and recorded with a vector network analyzer (see Method and Supplementary Note 3). Figure 3a shows a top view of the fabricated photonic crystal without the top aluminum plate, where the three cylinders at the corners are modified to a reduced radius of $r_1 = 1.4$ mm. To suppress spurious resonances between the parallel metal plates, the surrounding region is filled with microwave-absorbing sponges.

The numerically calculated bulk band structure of the designed photonic crystal (Fig. 3b) exhibits a trivial bandgap between the second and third bands. Figure 3c presents the calculated eigenfrequencies of the finite triangular structure, clearly distinguishing bulk modes from corner-localized states residing inside the bandgap. To experimentally identify these corner states, the sample is illuminated by a plane wave incident from the $+y$ direction, and transmission spectra are measured at two representative locations: the bulk center and the top corner of the structure. The corresponding spectrum of green (bulk) curve reveals a pronounced bandgap spanning from 6.89 to 8.83 GHz (Fig. 3d), evidenced by a sharp suppression of bulk transmission. This frequency range is consistent with the numerical predictions (Figs. 3b-3c).

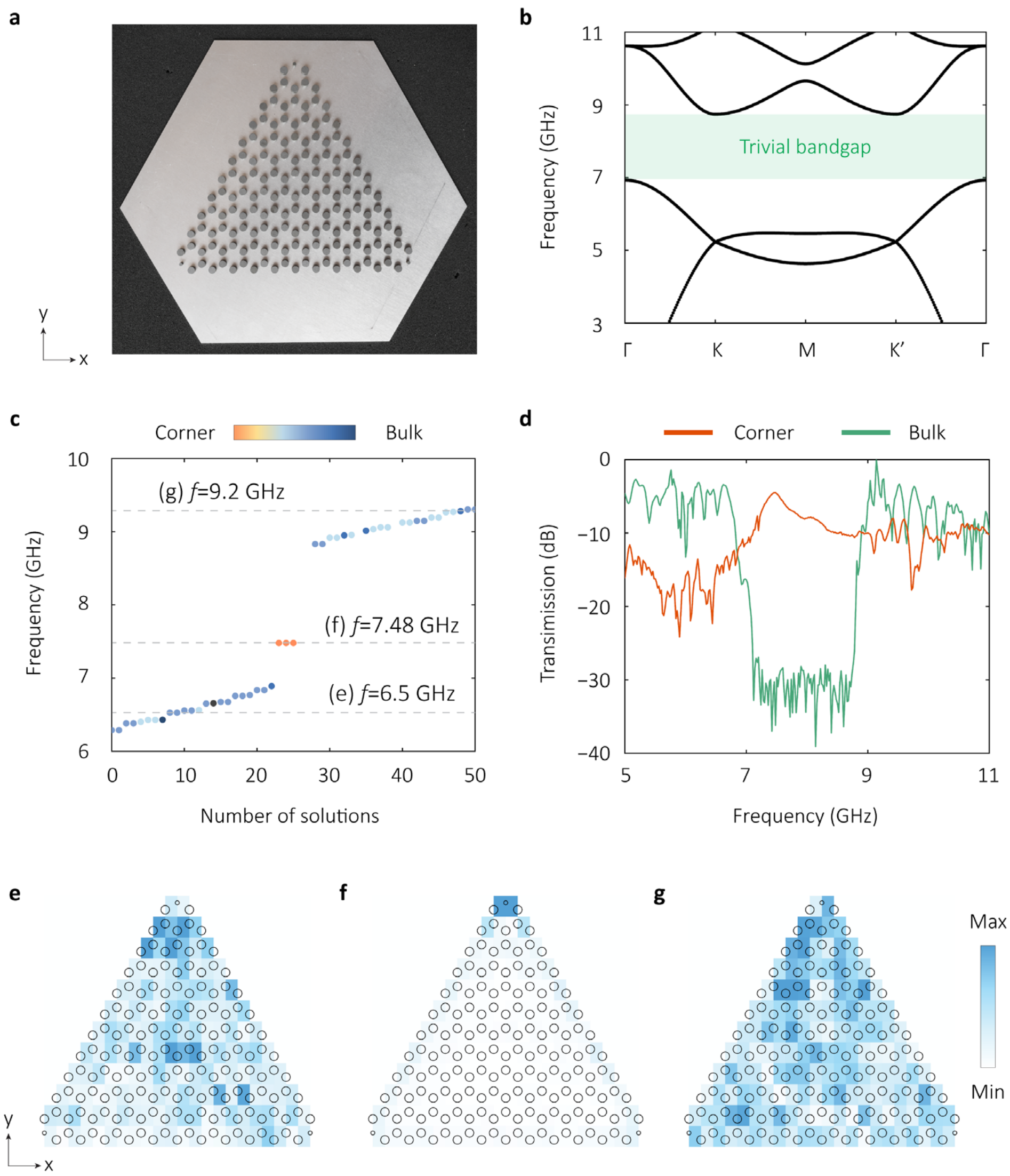

**Fig. 3. Observation of localized corner states.** (a) Top view of the fabricated triangular photonic crystal without the top aluminum plate. (b) Numerically calculated bulk band showing a trivial bandgap (shaded region). (c) Calculated eigenfrequencies of bulk and corner states for the finite triangular structure. (d) Measured transmission spectra at bulk (green) and corner (orange) positions under plane-wave excitation. (e-g) Measured electric-field distributions at representative frequencies: (e) $f$ = 6.5 GHz (bulk state), (f) $f$ = 7.48 GHz (corner state), and (g) $f$ = 9.2 GHz (bulk state).

Notably, a prominent transmission peak emerges at 7.48 GHz in the corner spectrum (orange curve in Fig. 3d), signaling the excitation of a localized corner mode. Near-field scanning at this frequency confirms strong field confinement at the top corner (Fig. 3f). For comparison, Figs. 3e and 3g display the measured electric field distributions corresponding to bulk modes in the lower and upper bands, respectively. Minor discrepancies between simulated and measured frequencies are attributed to fabrication tolerances in the YIG cylinder dimensions. We also note that the two bottom corners are not efficiently excited under plane wave illumination, owing to the absence of bulk or edge transport channels within the bandgap that could deliver energy to these corners. Overall, the experimental observations show good agreement with numerical simulations and unambiguously confirm the existence of programmable corners states.

Beyond this trivial-gap realization, we further demonstrate that the same local geometry engineering strategy can be extended to a topological Kagome lattice (see Supplementary Note 5), where tailored corner states are likewise constructed and observed through combined numerical design and experimental measurements. These results validate our central conclusion: in our scheme, corner-state formation is governed by local tailoring of flatband edge states, rather than by global symmetry breaking or higher-order bulk topological invariants.

**Robust corner states.** We further examine the robustness of the resulting states against structural perturbations, focusing on two representative types of disorder: position disorder and radius disorder[54–58]. Figures 4a-4c illustrate the unperturbed lattice together with the two disordered configurations. To quantify the disorder strength, we introduce a dimensionless fluctuation parameter $\delta$. For position disorder (Fig. 4b), $\delta$ denotes the maximum displacement $\Delta\mathbf{x}$ of a cylinder center from its original position, defined by $|\Delta\mathbf{x}| = \delta a$. For radius disorder (Fig. 4c), $\delta$ characterizes the maximum deviation $\Delta r$ from the nominal cylinder radius, defined as $|\Delta r| = \delta r$. In all simulations, the photonic crystal consists of dielectric cylinders with radius $r = 0.176a$ and relative permittivity $\varepsilon = 11.5$, while the corner cylinder is locally modified to a reduced radius of $r_E = 0.45r$.

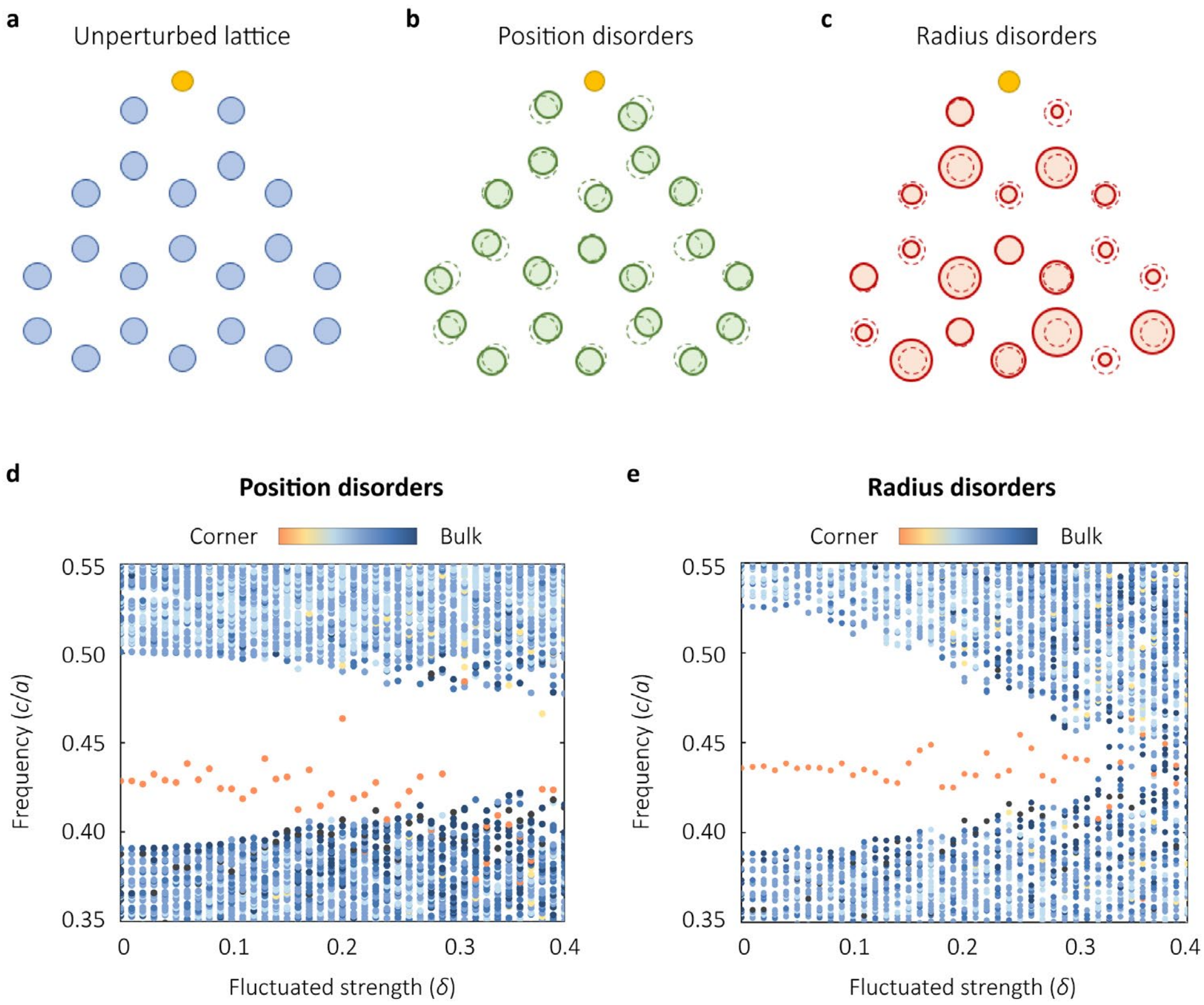

**Fig. 4. Robustness of corner states under structural disorder.** (a)-(c) Schematic diagrams of photonic crystal configurations: (a) Without perturbation. (b) Position disorders. (c) Radius disorders. (d)-(e) Eigenfrequencies of corner and bulk modes as a function of disorder strength for (d) position disorders and (e) radius disorders.

Figures 4d and 4e present the calculated eigenfrequencies of bulk and corner modes as functions of the disorder strength $\delta$ for position and radius perturbations, respectively. In both cases, the bandgap and the corner-localized state remain clearly identifiable at low disorder levels. For position disorder (Fig. 4d), the bandgap persists even under relatively strong perturbations, with the corner state merging into the bulk continuum only when $\delta$ approaches 0.25. In contrast, for radius disorder (Fig. 4e), the bandgap closes at $\delta \approx 0.37$, and the corner state becomes indistinguishable from bulk modes when $\delta$ exceeds approximately 0.31. This comparison indicates that the bandgap is considerably more sensitive to radius fluctuations than to position displacements.

These results demonstrate that the corner states survive moderate levels of structural disorder, owing to the stability of the bandgap supported by intrinsic long-range structural order[55,59]. At sufficiently high disorder strengths, however, the long-range order of the photonic crystal is progressively destroyed, leading to the eventual loss of corner-state localization as the modes merge into the bulk spectrum. Notably, corner states protected by global symmetries are typically far more vulnerable to position disorder, since their formation relies on delicate interference conditions that are easily disrupted by random displacements. The disorder sensitivity of such symmetry-dependent corner states in topological Kagome lattice is illustrated in Supplementary Note 5. By contrast, the corner states realized here, derived from flatband edge states and not reliant on global symmetry protection, exhibit enhanced robustness against position perturbations.

**Discussion and Conclusion.**

In summary, we have proposed and experimentally demonstrated a new class of robust corner states enabled by tailoring the flatband edge states originating from paired Dirac points. Rather than relying on global lattice-symmetry breaking, higher-order bulk topological invariants, or conventional topological bandgap engineering, our approach exploits local control of the boundary on-site mass in a honeycomb photonic crystal. This local tailoring deterministically shifts the flatband edge state across the upper bulk band and relocates it into a fully trivial bandgap, thereby inducing edge and corner localization. By modifying only the corner cylinders, we achieve spectra tunability and deterministic programmability of both the number and eigenfrequencies of corner states, without altering the global lattice geometry or bulk properties. Disorder analysis further confirms that these corner states remain robust under moderate structural perturbations and exhibit enhanced tolerance to positional disorder.

Together, these results establish boundary-tailored flatband edge state as a new pathway for generating and programming corner states in Hermitian platforms. Owing to the scalability of photonic crystal platforms and the all-dielectric nature of our design, these robust corner states can be readily extended to the optical regime via standard nanofabrication techniques, providing a practical route toward on-chip, disorder-resilient photonic devices.

## Methods

**Numerical methods.** The tight-binding model calculations in Figs. 1d-1g were performed in MATLAB using a two-dimensional honeycomb ribbon with $N = 11$ unit cells along the finite direction. The nearest-neighbor hopping amplitudes were set to $t_1$ and the boundary sites had an additional mass $M_s$. The surface density of states of the edge flatbands was obtained by calculating projected bandstructures and integrating eigenstates over the topmost and bottommost sites (see Supplementary Note 1). The corner states were calculated in a triangular geometry, where additional mass $M_s$ was assigned to the topmost site. For the surface density of states associated with electromagnetic projected bandstructures, supercell structures were first constructed in COMSOL. The resulting eigenmatrices were then exported to MATLAB, where the transfer matrix method was utilized to calculate the surface density of states (see Supplementary Note 2 and Ref. 53). The optical modes and electromagnetic bandstructures were calculated by finite-element-method solver in COMSOL Multiphysics (see Supplementary Note 3 for the detailed boundary settings).

**Microwave measurements.** In our experimental setup, a vector network analyzer (Keysight P9374A) was used to excite and probe electromagnetic waves. For plane-wave incidence, the excitation port was connected to a horn antenna, while a monopole antenna was used for measurement. The spectra were measured from 5 GHz to 12 GHz with a resolution of 0.02 GHz.

For the transmission spectra measured at the bulk and corner positions in Fig. 3d, the sample was illuminated by a horn antenna from the $+y$ direction. A monopole probe antenna was inserted through the predrilled holes in the aluminum plate and positioned at the selected bulk and corner sites to detect the local electromagnetic field (see Supplementary Note 3). To obtain the electric-field distributions displayed in Figs. 3e–3g, we scanned the whole sample area in the $x$-$y$ plane through the reserved holes one by one. The electromagnetic field at specific frequencies was then reconstructed.

## Acknowledgements

We acknowledge the financial support from the Science and Technology Project of Guangdong (2020B010190001), and the National Natural Science Foundation of China (12434016).

**Author contributions**

Z. J., J. C., and Z.-Y. L. conceived and designed this research project. Z. J. and J. C. performed the tight-binding calculations and electromagnetic simulations. Z. J. and Y. Z. fabricated the samples and conducted the experiments. Z. J., J. C., Z. D., and Z.-Y. L. analyzed the data and wrote the manuscript. J. C., Z. D., and Z.-Y. L. supervised the project.

**Competing interests**

The authors declare no competing interests.

*Supplementary Information*

# Observation of robust corner states in photonic crystals without global symmetries

Zitao Ji[1,2,#], Jianfeng Chen[3,#,*], Yidong Zheng[1], Zhaogang Dong[2,*], and Zhi-Yuan Li[1,*]

[1] School of Physics and Optoelectronics, South China University of Technology, Guangzhou 510640, China

[2] Science, Mathematics, and Technology (SMT), Singapore University of Technology and Design, 8 Somapah Road, 487372, Singapore

[3] State Key Laboratory for Mesoscopic Physics and Frontiers Science Center for Nano-optoelectronics, School of Physics, Peking University, Beijing 100871, China

[*] Correspondence: phzyli@scut.edu.cn; cjf@pku.edu.cn; zhaogang_dong@sutd.edu.sg

[#] These authors contributed equally to this work.

**Contents**

**Supplementary Note 1: Analytical calculation of tight-binding model.**

We consider a honeycomb lattice ribbon with lattice constant $a = 1$, consisting of two sublattices A and B (Fig. S1a). The wave vector is $\mathbf{k} = (k_x, k_y)$, and the lattice vectors are $\boldsymbol{\delta}_1 = (0, \frac{\sqrt{3}}{3})$, $\boldsymbol{\delta}_2 = (-\frac{1}{2}, -\frac{\sqrt{3}}{6})$, and $\boldsymbol{\delta}_3 = (\frac{1}{2}, \frac{\sqrt{3}}{2})$, respectively. We set the nearest-neighbor couplings as $t_1$, and the parity-inversion breaking on-site potential as $M$. Therefore, the $k$-space Hamiltonian of Dirac semimetal is given by $H(1,1) = M$, $H(1,2) = t_1(e^{i\mathbf{k}\cdot\boldsymbol{\delta}_1} + e^{i\mathbf{k}\cdot\boldsymbol{\delta}_2} + e^{i\mathbf{k}\cdot\boldsymbol{\delta}_3})$, $H(2,1) = H^*(1,2)$, and $H(2,2) = -M$. To simplify the analysis, we consider the Hamiltonian along the $k_x$ direction and set $k_y = 0$. The Hamiltonian can be rewritten as $\mathbf{H} = \begin{bmatrix} M & t_1 + 2t_1\cos(\frac{k_x}{2}) \\ t_1 + 2t_1\cos(\frac{k_x}{2}) & -M \end{bmatrix}$. Solving the eigenvalue equation $H\psi = E\psi$ will yield the band energies as $E(k_x) = \pm\sqrt{(t_1 + 2t_1\cos(\frac{k_x}{2}))^2 + M^2}$, where $\psi$ denotes the eigenstate wave function. In our configuration, we set $M = 0$ and the energies at the K and K' valley will be $E_{\mathrm{k}} = E_{\mathrm{k'}} = 0$.

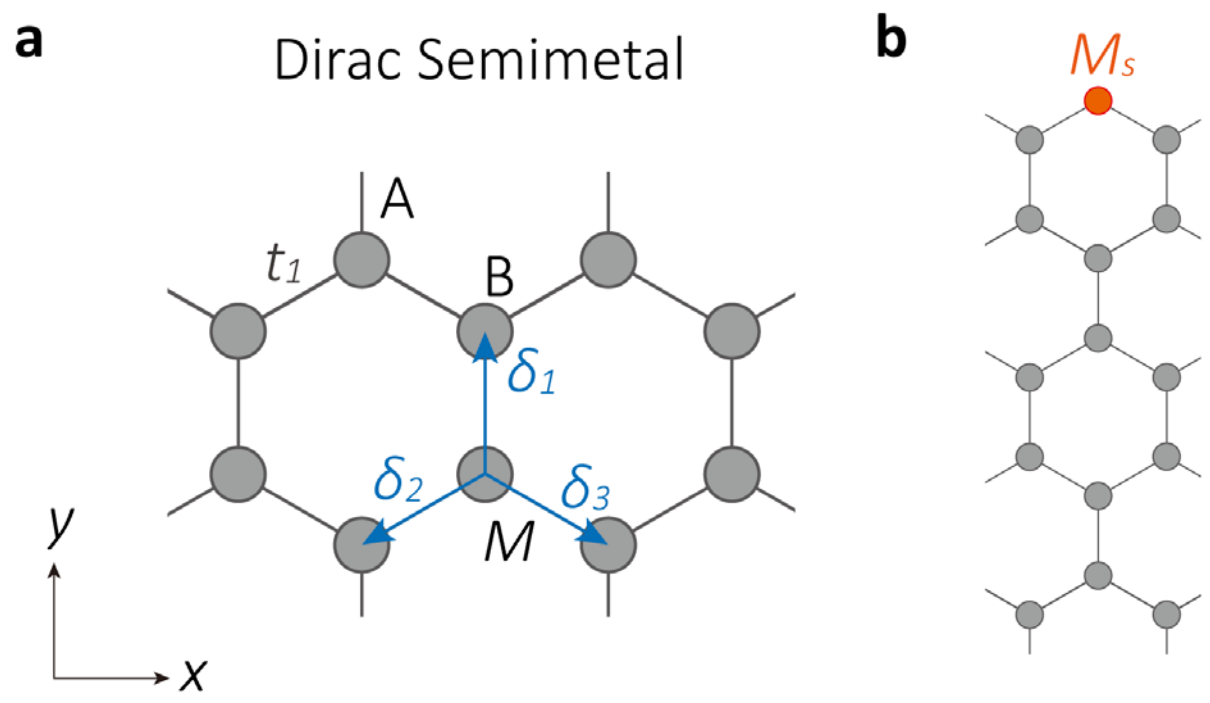

**Fig. S1. Tight-binding models.** (a) Dirac semimetal. The black solid lines represent the nearest-neighbor couplings. (b) Schematic for calculating projected bandstructure of a ribbon geometry. Periodic boundary conditions are applied along $x$ direction and the outermost boundary site has an additional on-site mass $M_s$.

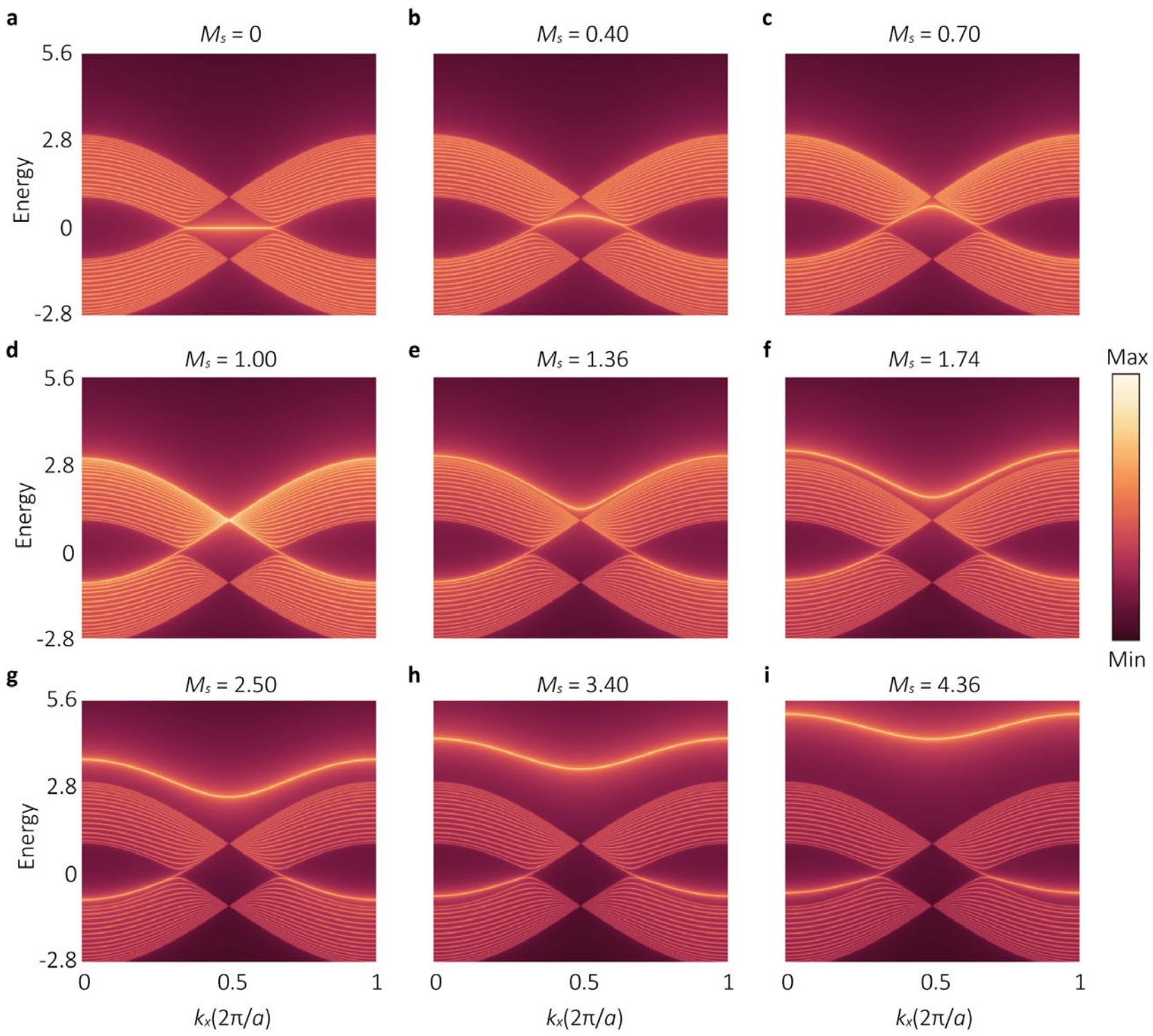

**Fig. S2. Projected bandstructures of the Dirac semimetal with different $M_s$.** The color scale indicates the surface density of states (SDOS), highlighting edge-localized modes. (a) $M_s = 0$. (b) $M_s = 0.40$. (c) $M_s = 0.70$. (d) $M_s = 1.00$. (e) $M_s = 1.36$. (f) $M_s = 1.74$. (g) $M_s = 2.50$. (h) $M_s = 3.40$. (i) $M_s = 4.36$.

Next, we consider a zigzag strip comprising $N$ cells in the $y$ direction and periodic boundary conditions along the $x$ direction, so that $k_x$ remains a good quantum number for calculating the project band (Fig. S1b). To implement flatband modification, we introduce an additional on-site mass $M_s$ only on the topmost and the bottommost sites of the ribbon. For example, the topmost site's Hamiltonian becomes $H(1,1) = M + M_s$. Figure1d-1e and Figure S2 show the projected bandstructure of a zigzag ribbon with $N$ = 11 cells for different values of $M_s$. To distinguish and highlight the edge states from the bulk states, we evaluate the SDOS. Specifically, we define the boundary weight by integrating the eigenstates $\psi_n$ over topmost and bottommost sites, which has the form $W_n = \sum_{i \in \text{edge}} \left|\psi_{n,i}\right|^2$. For bulk states, the wave functions extend across the entire ribbon, so the boundary contribution is small, yielding a small $W_n$. In contrast, edge states are concentrated near the ribbon boundaries, resulting in a larger $W_n$, which allows them to be highlighted in the projected spectrum. When $M_s$ = 0, a pair of flatbands appears between two Dirac points at $E = 0$ (Fig. S2a). As $M_s$ increases, the flatbands firstly shift up into the bulk bands, and then move into the trivial bandgap (Fig. S2b-S2i).

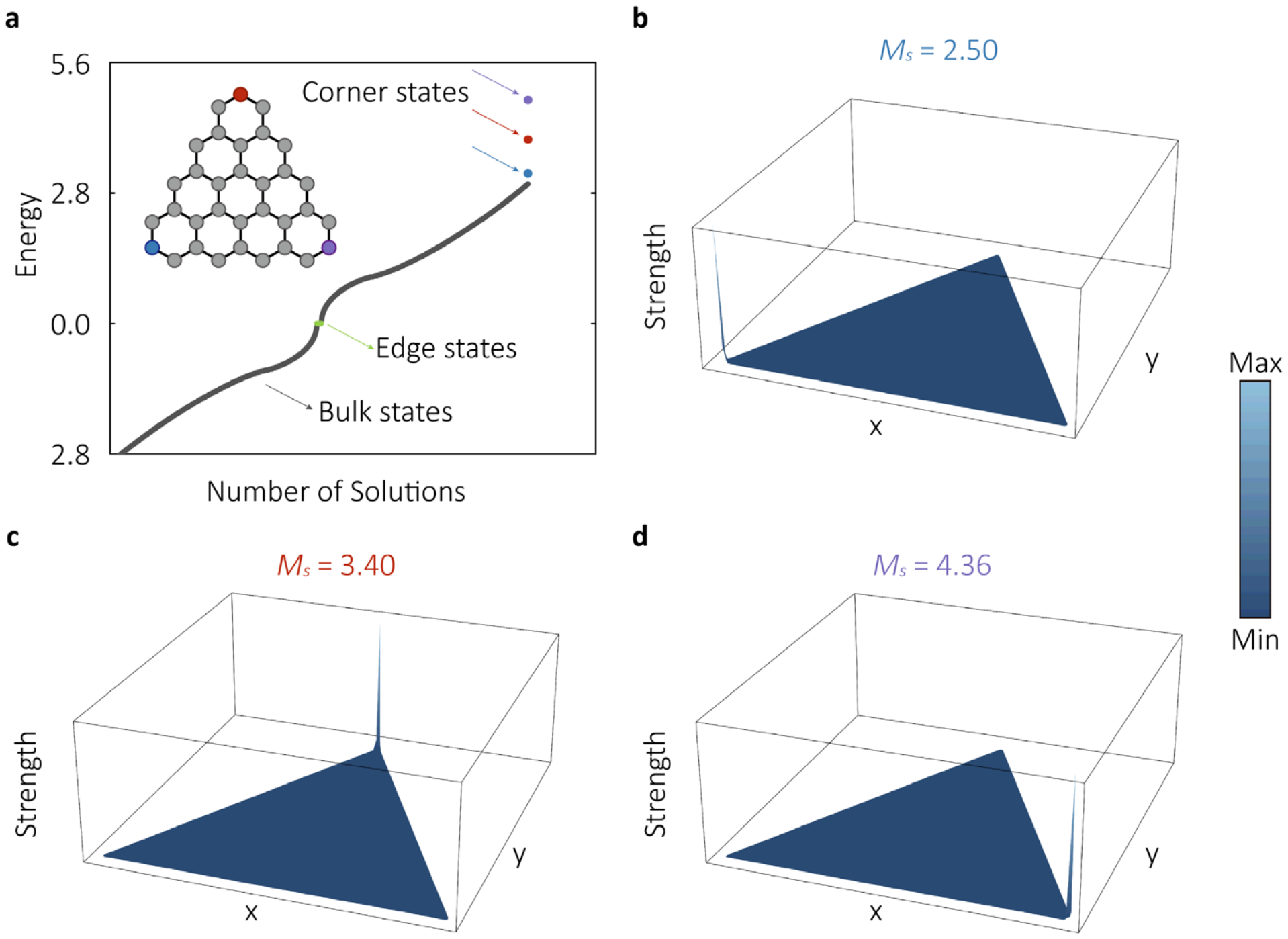

**Fig. S3. Programmable corner states in a triangular geometry with locally tailored additional on-site masses $M_s$.** (a) The bulk, edge, and topological corner eigenstates calculated in a triangular geometry. The leftmost, topmost, and rightmost sites have different additional on-site masses $M_s$. (b)-(d) Corresponding eigenstate field profiles showing corner localization for three distinct $M_s$. (b) $M_s$ = 2.50. (c) $M_s$ = 3.40. (d) $M_s$ = 4.36.

To calculate the corner states, we consider a finite triangular geometry where the leftmost, topmost, and rightmost sites are assigned different additional on-site masses $M_s$, chosen as 2.50, 3.40, and 4.36, respectively. The resulting spectrum is summarized in Fig. S3a and three in-gap corner states are clearly resolved. Figures S3b-S3d show the eigenstate fields of these three corner states, confirming that each mode is strongly localized at its designated corner while appearing at a distinct eigenenergy. This demonstrates corner states with programmable spectral positions enabled by local mass tailoring.

**Supplementary Note 2: Surface density of states of photonic crystals.**

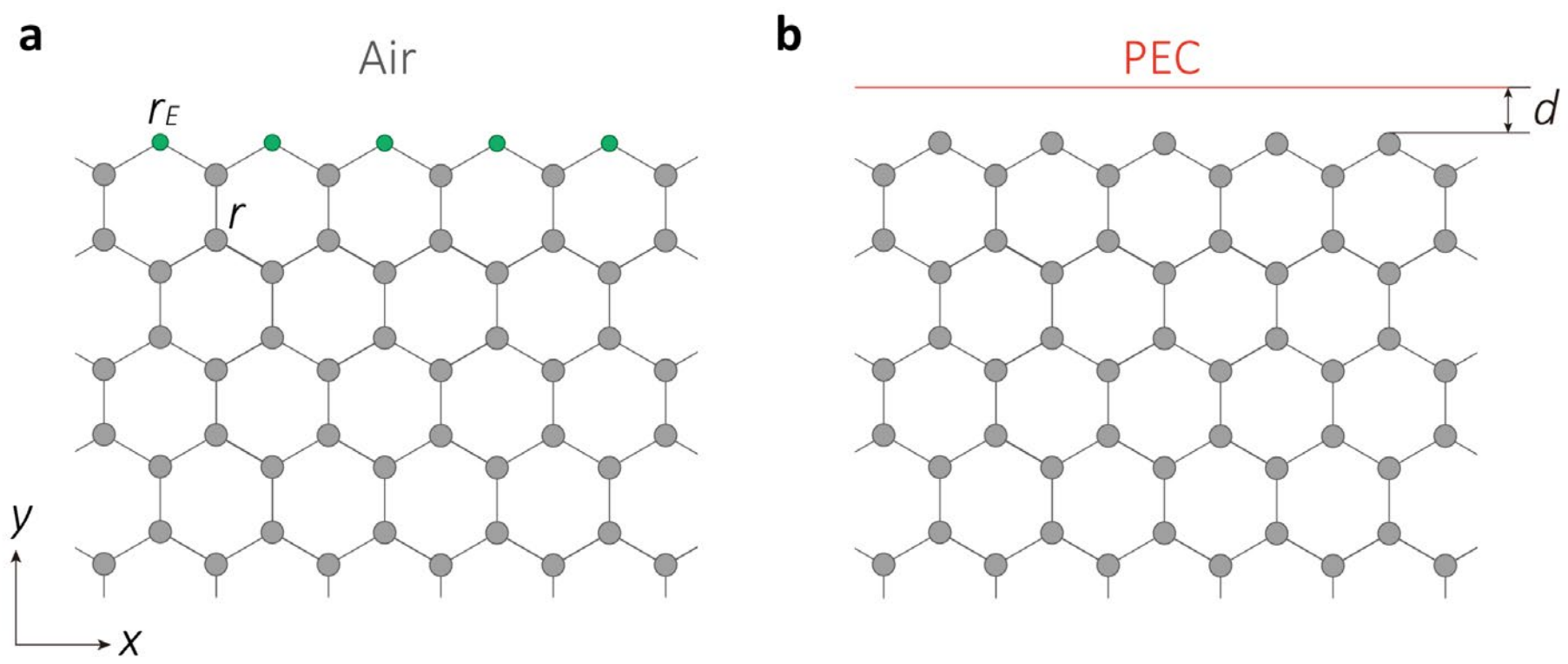

**Fig. S4. The surface configurations in photonic crystals.** (a) Semi-infinite photonic crystals with surface modified rods and an open boundary (air cladding). (b) Semi-infinite photonic crystals without rod modification, terminated by a PEC boundary placed at a distance *d* from the photonic crystal surface.

The SDOS calculations of photonic crystals are based on the approaches in Ref. 1. To obtain the SDOS spectrum associated with a well-defined single boundary of a semi-infinite system, we start from the surface Green's function:

$$[\nabla\times\mu^{-1}(\mathbf{r})\cdot\nabla\times-\omega^2\varepsilon(\mathbf{r})]G(\mathbf{r},\mathbf{r}';\omega)=I\delta(\mathbf{r}-\mathbf{r}') \qquad \text{(S1)}$$

where $\mu(\mathbf{r})$ and $\varepsilon(\mathbf{r})$ are the permeability and permittivity at location **r**, $I$ is a 3×3 identity tensor, and $\delta(\mathbf{r}-\mathbf{r}')$ is the Dirac delta source at $\mathbf{r}'$. Equation (S1) can be compactly written as $ZG = I$, with the operator $Z=\nabla\times\mu^{-1}(\mathbf{r})\cdot\nabla\times-\omega^2\varepsilon(\mathbf{r})$. For a semi-infinite periodic crystal, the system can be discretized into identical layers normal to the surface. Considering the intra-coupling and inter-coupling between neighboring layers, the operator $Z$ can be transformed into the following block-tridiagonal form:

$$Z = \begin{pmatrix} Z_{0,0} & Z_{0,1} & & \\ Z_{1,0} & Z_{0,0} & Z_{0,1} & \\ & \ddots & \ddots & \ddots \end{pmatrix} \tag{S2}$$

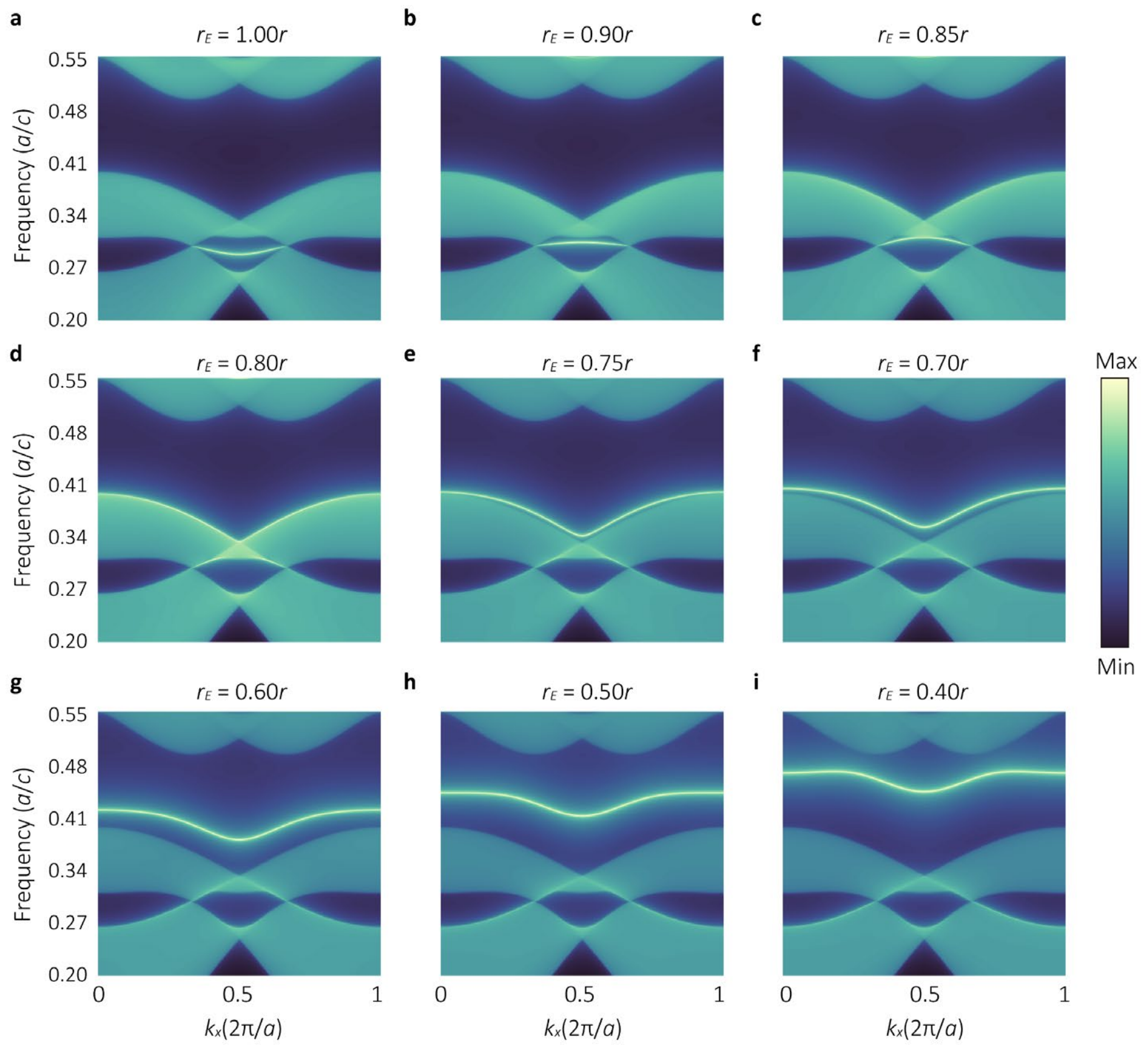

**Fig. S5. The SDOS calculations of the projected bandstructures of photonic crystals with different $r_E$.** (a) $r_E = 1.00r$. (b) $r_E = 0.90r$. (c) $r_E = 0.85r$. (d) $r_E = 0.80r$. (e) $r_E = 0.75r$. (f) $r_E = 0.70r$. (g) $r_E = 0.60r$. (h) $r_E = 0.50r$. (i) $r_E = 0.40r$.

Figure S4a illustrates the surface configuration of a semi-infinite photonic crystal, where the edge rods are locally tailored and exposed to air (open boundary). The surface Green's function can be derived through transfer matrix method[2] applied to equation (S2). The SDOS is then evaluated from the imaginary part of $G$ by introducing an infinitesimal imaginary frequency $\eta$ in the following equation:

$$\mathrm{SDOS}(r;\omega) = \frac{2\omega}{\pi}\mathrm{Im}\left\{\mathrm{Tr}\left[\varepsilon(\mathbf{r})\lim_{\eta\to 0^+} G(\mathbf{r},\mathbf{r};\omega+i\eta)\right]\right\} \tag{S3}$$

We calculate SDOS of projected bandstructures for different edge rod radii $r_E$ (Fig.

2 and Fig. S5). As $r_E$ is reduced, the edge bands will firstly move into the upper bulk bands (Fig. S5a-S5d), and become spectrally isolated inside the trivial bandgap between the second and third bands (Fig. S5e-S5i). In addition to geometric tailoring the edge rods, we find that the edge bands can also be tuned by modifying the separation between the photonic crystal and a perfect electric conductor (PEC) boundary (Fig. S4b). The edge bands can also shift into the trivial bandgap by reducing the distance $d$ (Fig. S8).

**Supplementary Note 3: The configurations for numerical calculation of corner states and experimental setups.**

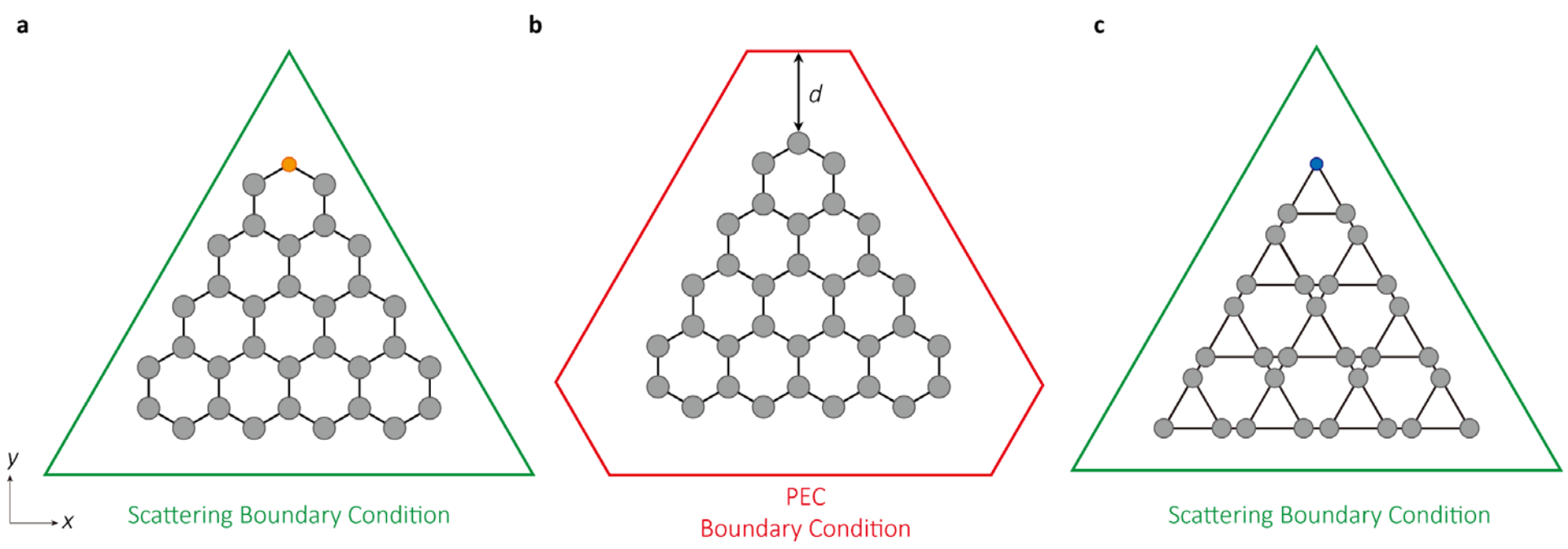

**Fig. S6. The configurations for numerical calculations of corner states.** (a) The configuration with modified corner rods in a honeycomb photonic crystal, surrounded by scattering boundaries. (b) Honeycomb photonic crystal enclosed by PEC boundaries. (c) The configuration with modified corner rods in a Kagome photonic crystal, surrounded by scattering boundaries.

The corner states in Fig. 2-4 are calculated by the configuration in Fig. S6a. In this setup, a finite triangular honeycomb photonic crystal is embedded in an unbounded $x$-$y$ domain and enclosed by scattering boundaries to emulate an open system and suppress spurious reflections.

In Supplementary Note 4, the corner states are instead calculated in the PEC confined geometry illustrated in Fig. S6b, where the honeycomb photonic crystal is placed within PEC boundaries. The separation between the PEC boundaries and the unmodified photonic crystal surface is denoted by $d$.

For the Kagome-lattice corner states discussed in Supplementary Note 5, we again adopt an open-system configuration (Fig. S6c): a finite Kagome photonic crystal with locally modified corner rods is surrounded by scattering boundaries in the *x*-*y* plane.

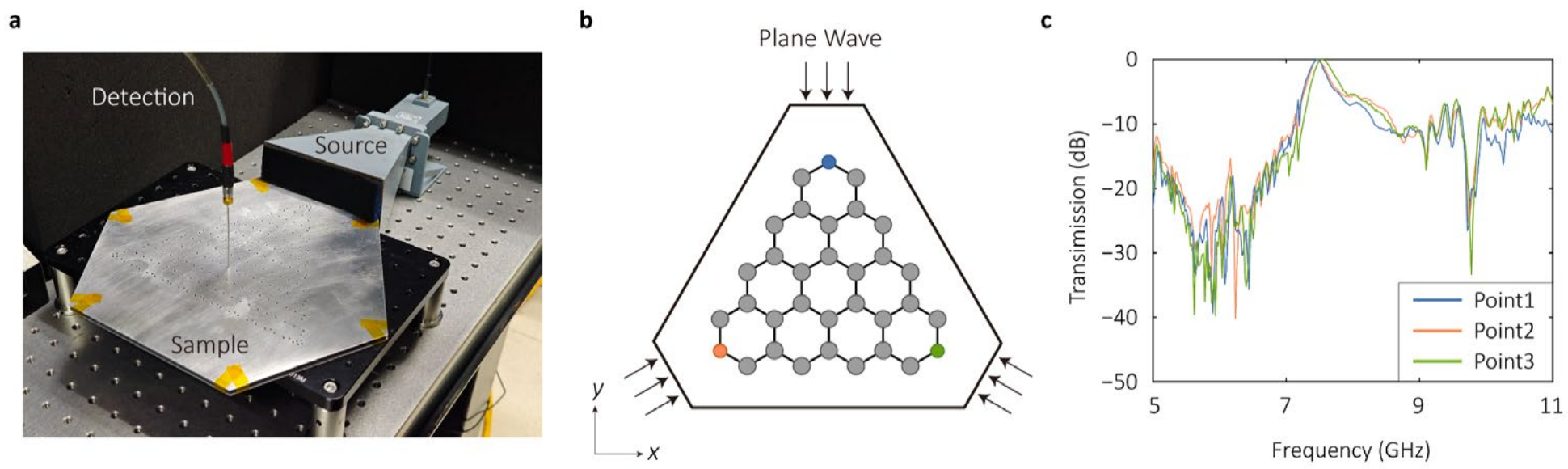

**Fig. S7. Experimental setup and transmission measurements of corner states.** (a) Overall picture of the measuring process. (b) Schematic of the measurement configuration, with plane-wave excitation applied from three sides. (c) Transmission spectra measured at the three corner positions (blue, orange, and green).

The experimental setup is displayed in Fig. S7a. To verify fabrication consistency, we measure the transmission spectra by exciting the sample from three different sides and probing the corresponding corners (Fig. S7b-S7c). In all cases, the resonance peaks appear at nearly the same frequency, with variation within the fabrication tolerance of the YIG cylinders (±0.1mm) and measurement uncertainty.

**Supplementary Note 4: Modification of corner states through PEC boundaries.**

In this section, we show that instead of modifying the radii of the corner rods, reducing the PEC boundary spacing can also shift Dirac flatbands into the trivial bandgap between the second and third bands. The corresponding geometry configuration is illustrated in Fig. S4b, where the distance between the photonic crystal edge and the PEC wall is denoted by *d*. We calculate the SDOS of projected bandstructures in photonic crystals for progressively reduced *d* (Figs. S8a-S8f). As the PEC walls approach the photonic crystal, the two edge bands between the Dirac points firstly move upward into the bulk bands (Fig. S8d). When *d* further decreases, these edge bands shift into the bandgap region (Fig. S8e-S8f). Therefore, isolated corner states can also be constructed within the bandgap purely by compressing the PEC boundary spacing, without any geometric modification of the dielectric rods (Fig. S8g and S8h). The eigenstate result confirms the emergence of corner modes within the bandgap, separated from the bulk states (Fig. S8g). The corresponding electromagnetic field distribution in Fig. S8h further demonstrates strong localization at three corner rods, showing that PEC spacing alone can be used to engineer corner states through flatband shifting.

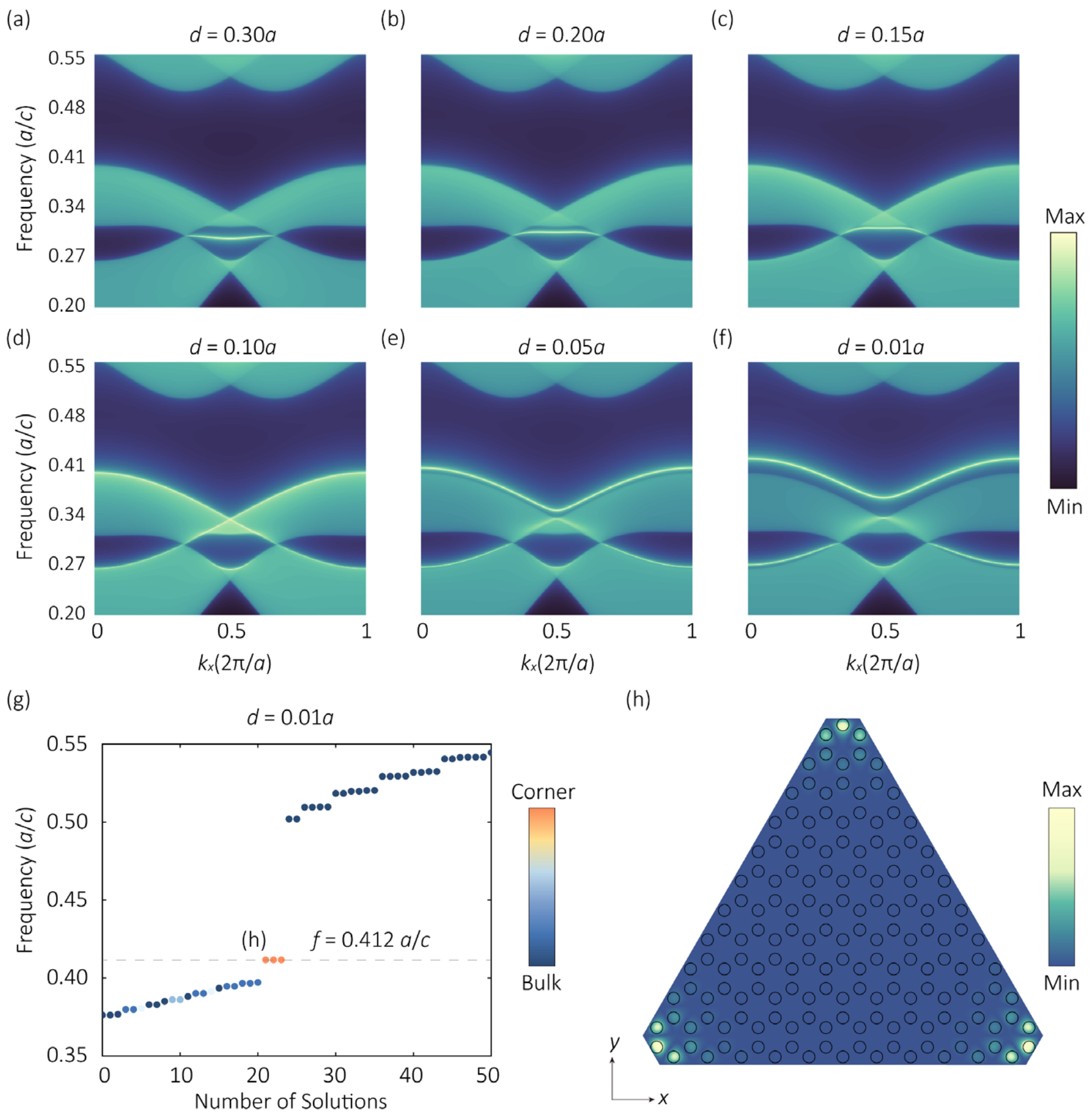

**Fig. S8. Modification of corner states by varying the PEC boundary separation.** (a)-(f) The SDOS calculations of projected bandstructures with different distance $d$ between the edge of photonic crystals and the PEC boundaries. (a) $d$ = 0.3$a$. (b) $d$ = 0.2$a$. (c) $d$ = 0.15$a$. (d) $d$ = 0.10$a$. (e) $d$ = 0.05$a$. (f) $d$ = 0.01$a$. (g) The eigenmodes of the bulk (blue dots) and corner modes (orange dots) in the triangle photonic crystal

structure when $d = 0.01a$. (h) The electric field distribution of the corner modes at $f = 0.412a/c$, showing strong localization near the three corner rods.

This boundary-driven spectral restructuring differs fundamentally from both conventional defect-induced localization and previously reported topological corner states, which rely on either local perturbations or bulk topological invariants and symmetry protection. To make this distinction explicit, Table S1 systematically compares conventional defect modes, higher-order topological corner states, and the present mechanism in terms of physical origin and required conditions. This comparison highlights that the reported states constitute a complementary route to robust localization beyond existing paradigms.

**Table S1 | The configuration and mechanism difference between traditional defect modes, topological corner states, and our work**

| | Defect mode | Topological corner state | **Our work** |
|---|---|---|---|
| Configuration | Modification single sites within the structure 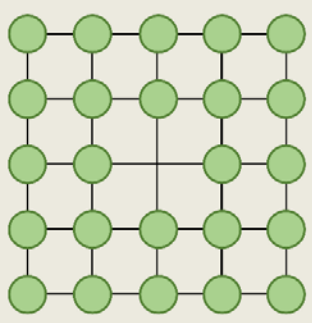 | Shrunken-expanded lattices or breathing Kagome lattices. 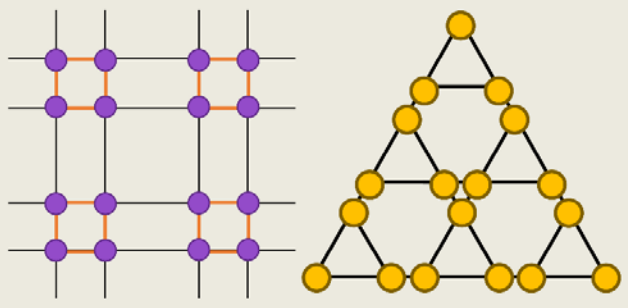 | Boundary engineering of single sites or boundary conditions. 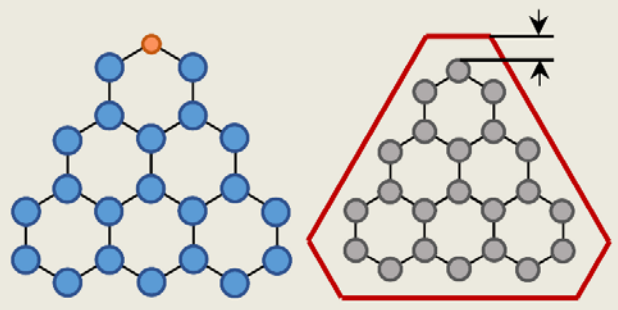 |
| Mechanism | Defect mode within the bandgap due to Bragg reflection. | Higher-order topological state within an opened topological bandgap due to symmetry breaking. | The relocating of Dirac flatbands into trivial bandgap without opening a topological bandgap. |

**Supplementary Note 5: The design of tailored corner states in breathing Kagome lattice.**

**(i) Photonic design.**

In this section, we study the locally geometric tailored corner states in a breathing Kagome lattice case. The Kagome structure is arranged in equilateral triangle structure (Fig. S9a). Each unit Kagome cell consists of three rods with radius $r = 0.1a$. The coupling strength between the three rods within the same cell is denoted as $t_1$ and the coupling strength between rods in adjacent cells is set as $t_2$. To open a bandgap, we vary the coupling strengths $t_1$ and $t_2$ by shifting the cylinders closer to or farther away from one another. For the unperturbed Kagome lattice, the distance between the three rods in one unit cell is set as $d = 0.5a$ consistently (Fig. S9b, position 1), which means that inter-cell and intra-cell coupling strengths $t_1$ and $t_2$ are identical. The bandstructure in this case exhibits two degenerate Dirac points at K and K' points in the Brillouin zone, as indicated by the blue lines in the bulk bandstructures in Fig. S9c. When the distant $d$ is tuned to be larger or smaller than $0.5a$, corresponding to positions 2 and 3 in Fig. S9b, the crystalline symmetry is reduced, leading to the opening of a complete topological bandgap. Note that, topologically nontrivial structures are corresponded with expanded structure where $d > 0.5a$, for which $|t_1/t_2| < 1$. By contrast, the shrunken structures with $d < 0.5a$ correspond to topologically trivial structures. The topological nontrivial expanded structure and trivial shrunken structure have identical bulk bandstructures, indicated by the red lines in Fig. S9c.

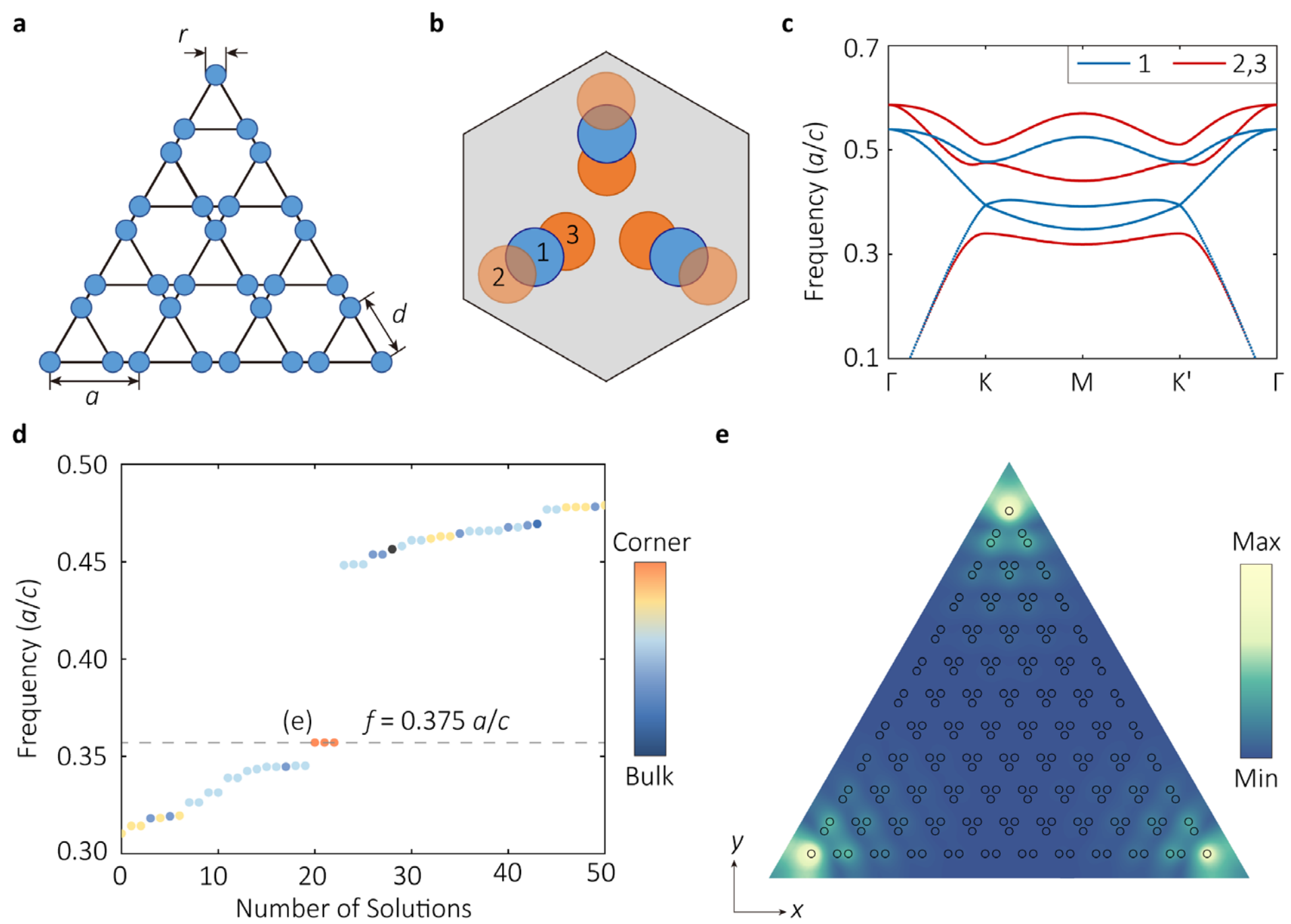

**Fig. S9 The construction of symmetry-dependent corner states in Kagome lattice.** (a) The schematic of the Kagome triangular photonic crystal. (b) The lattice unit cells of unperturbed (1), expanded (2), and shrunken (3) Kagome lattices. (c) The bulk

bandstructures of unperturbed (1), expanded (2), and shrunken (3) Kagome lattices. (d) The eigenfrequencies of modes in Kagome triangular structure. (e) Electric field distribution of symmetry-dependent corner states at the normalized frequency $f$ = 0.357($a/c$).

In our configuration, we set $d = 0.7a$ for opening a topological bandgap, and Fig. S9d shows the calculated eigenfrequencies in the normalized frequency range 0.30~0.50($a/c$). There will be Type-I, Type-II corner states, and edge states appeared within the bandgap (see Refs. 3,4). In our discussion, we focus on Type-I corner modes (highlight with dash lines in Fig. S9d). These three symmetry-dependent corner modes appear at three corners, as the field profile shown in Fig. S9e.

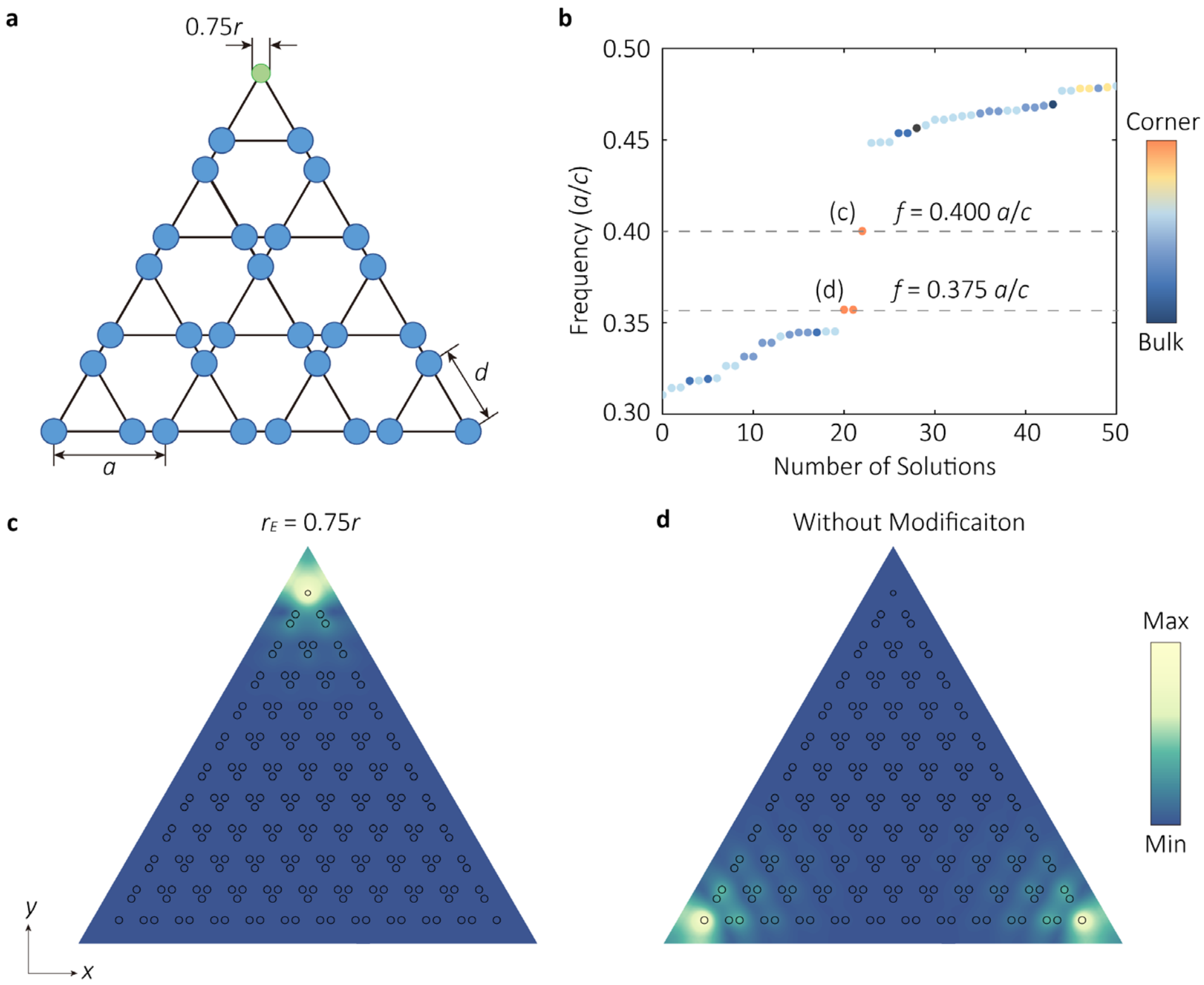

**Fig. S10. The construction of tailored corner states in Kagome lattice**. (a) Schematic diagram of the corner-tailored Kagome triangular photonic crystal. (b) Calculated eigenfrequencies of modes in Kagome triangular structure. (c)-(d) Electric field distributions of (c) symmetry-independent corner states at the normalized frequency $f$

= 0.400 ($a/c$), and (d) symmetry-dependent corner states at the normalized frequency $f$ = 0.357 ($a/c$).

Then, we set the topmost pillar radius to $r_E$ = 0.75$r$ (Fig. S10a). The calculated eigenfrequencies (Fig. S10b) indicate that the symmetry-independent corner state has an eigenfrequency of 0.400($c/a$). The unperturbed leftmost and rightmost rods still support the symmetry-dependent corner modes, which are located at lower frequency 0.357($c/a$) in Fig. S10b. Figures S10c and S10d show the electric field distribution of symmetry-independent corner mode and symmetry-dependent corner modes, respectively. Therefore, our approach of constructing corner state through local modification of the corner rod yields a different eigenfrequency and more localized field distribution compared to the symmetry-dependent ones. We have illustrated that such tailored topological corner states can be realized within both trivial and nontrivial bandgaps.

**(ii) Experimental identification.**

We investigate the symmetry-independent corner states in a Kagome triangular photonic crystal with a lattice constant of 15 mm. The radii of the YIG rods are 1.5 mm, while the corner cylinders are tailored by ones with radii of 1.125 mm (Fig. S11a). Figure S11b shows the numerically calculated eigenfrequencies of the corner and bulk modes. A bandgap can be observed approximate from 6.92 to 8.78 GHz, as indicated by the black line in Fig. S11c. The red line shows the transmission spectrum at the topmost corner of the Kagome triangle structure, where a peak at 8 GHz is observed. The electric field (Fig. S11e) indicates strong field confinement at the topmost corner. The slight discrepancy between the simulated and experimental field profiles arises from the fixed probe position used in the near-field scanning, together with the intrinsically broader spatial distribution of the Kagome corner state. For comparison, Figs. S11d and S11f show the field distributions of the bulk modes in the lower and upper bands. Hence, we experimentally observe symmetry-independent corner states within nontrivial bandgap here and trivial bandgap in the main text, showing that our approach of constructing tailored corner state can be applied in different photonic crystal systems.

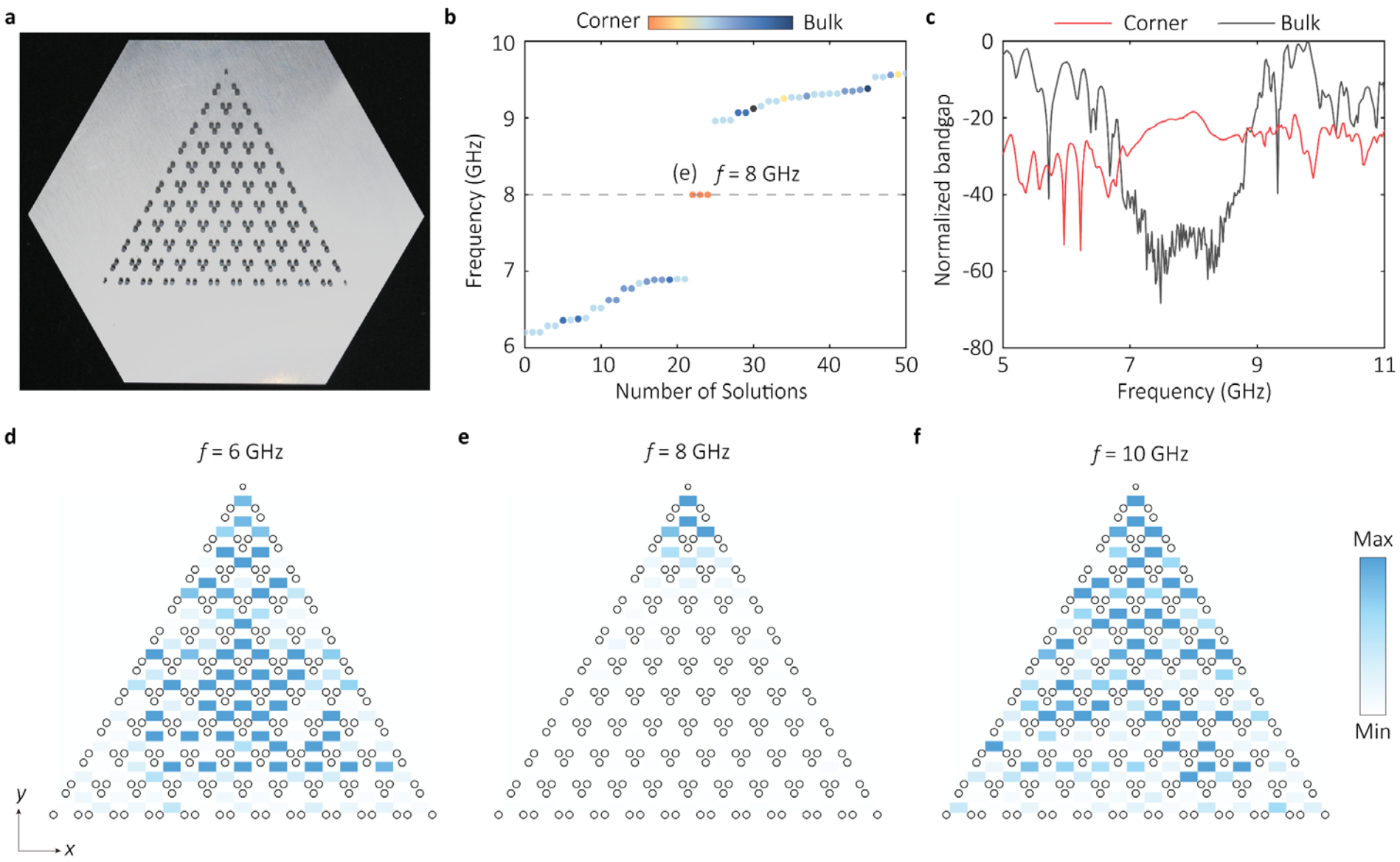

**Fig. S11. Experimental observation of tailored corner state in Kagome lattice.** (a) Top view of the Kagome triangular sample, without the top aluminium plates. (b) Calculated eigenfrequencies of bulk and corner states. (c) Measured transmission spectra at the bulk (black line) and corner (red line) positions under plane wave excitation. (d)-(f) Measured electric field distributions at representative frequencies: (d) 6 GHz (bulk mode), (e) 8 GHz (corner mode), and (f) 10 GHz (bulk mode).

**(iii) Corner states stability against disorders in Kagome lattices.**

We also investigate the stability of the corner states in breathing Kagome lattices. Figures S12a – S12c show the unperturbed Kagome lattice and two types of disordered configurations. We utilize the same fluctuated strength ($\delta$) definition in the main text for characterizing the degree of disorders. In all calculations, the photonic crystal is composed of dielectric cylinders with radii of $r = 0.1a$ and the relative permittivity $\varepsilon = 11.5$, while the modified corner rod is set to $r_E = 0.75r$.

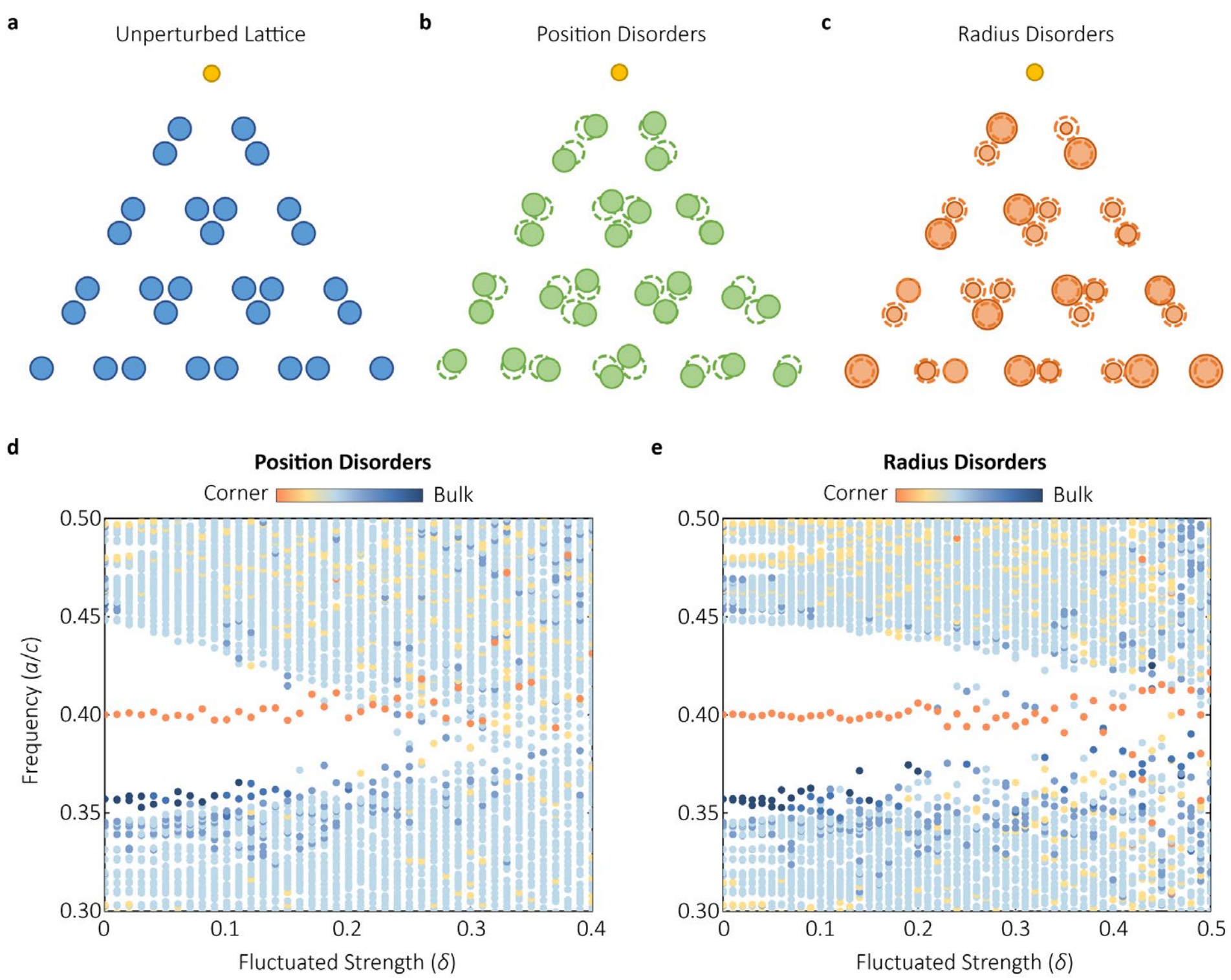

**Fig. S12. Robustness of corner states under structural disorder.** (a)-(c) Sketch maps of the Kagome photonic crystals: (a) Unperturbed Kagome photonic crystal. (b) Position-perturbation photonic crystal. (c) Radius-perturbation photonic crystal. (d)-(e) Calculated eigenfrequencies of corner and bulk modes as a function of disorder strength for (d) position disorder and (e) radius disorder.

Figures S12d and S12e show the eigenfrequencies in Kagome lattices as a function of $\delta$ for the position perturbations and radius perturbations, separately. In position disorder cases, the topological Kagome bandgap close more quickly than the trivial bandgap shown in Fig. 4d. The Kagome bandgap is closed when $\delta$ is approximately larger than 0.35, and the corner modes merge into bulk modes when $\delta > 0.20$. We note that the topological Kagome bandgap is protected by the global symmetries. Therefore, this bandgap is more affected by the position perturbations, since the symmetry is associated with specific interference pathways in the lattice. However, in radius disorder case, the topological bandgap still exists when $\delta$ is larger than 0.50 (Fig. S12e). Compared with trivial bandgap displayed in Fig. 4e, the topological bandgap is more stable in radius disorder case, possibly because this type of perturbations has less impact on the global symmetries. However, with the increase of the fluctuated strength, there will be other localized modes exist within the bandgap. The corner modes may merge into bulk modes when $\delta > 0.42$. Therefore, the corner states constructed in the main text — being independent of global symmetry — exhibit enhanced robustness against position perturbations.